\documentclass[twocolumn,english,prb,longbibliography,hypertext,superscriptaddress]{revtex4-2}
\usepackage[utf8]{inputenc}
\usepackage{amsmath}
\usepackage{amstext,amssymb}
\usepackage{amsthm}
\usepackage{graphics,graphicx}
\usepackage{verbatim}
\usepackage{color}
\usepackage{braket}
\usepackage{mathrsfs}
\RequirePackage[colorlinks=true,linkcolor=blue]{hyperref}
\usepackage{dsfont}
\usepackage{bm}
\usepackage{bbm}
\usepackage{siunitx}

\definecolor{gray}{rgb}{.6,.6,.6}
\definecolor{asparagus}{rgb}{0.53, 0.66, 0.42}

\newcommand{\ketbra}[2]{|{#1}\rangle\!\langle{#2}|}

\usepackage[normalem]{ulem}

\newcommand{\Eqref}[1]{Eq.~(\ref{#1})}                                

\newcommand{\Figref}[1]{Fig.~\ref{#1}}
\newcommand{\Secref}[1]{Sec.~\ref{#1}}
\newcommand{\Appref}[1]{App.~\ref{#1}}

\newcommand{\tr}[1]{\mathrm{Tr}[#1]}
\newcommand{\ie}{\emph{i.e.}}

\graphicspath{{Figures/}}

\newcommand{\dipc}{Donostia International Physics Center (DIPC), Manuel Lardizabal Pasealekua 4, E-20018 Donostia-San Sebastian, Spain}
\newcommand{\ikerbasque}{IKERBASQUE, Basque Foundation for Science, Euskadi Plaza 5, E-48009 Bilbao, Spain}

\begin{document}
\title{Creation and characterization of leviton excitations in tight-binding chains}

\author{Stephen R. McMillan}
\affiliation{\dipc}

\author{Thomas Frederiksen}
\affiliation{\dipc}
\affiliation{\ikerbasque}

\author{Géza Giedke}%
\affiliation{\dipc}
\affiliation{\ikerbasque}

\date{\today}

\begin{abstract}
Levitons are minimal-excitation electronic wave packets generated by Lorentzian voltage pulses and constitute a central resource for electron quantum optics. Their creation and defining properties are usually formulated in continuum scattering descriptions, whereas many candidate platforms for integrated electronic quantum circuits---including quantum-dot arrays, graphene nanoribbons, and moir\'e materials---are finite, discrete, and strongly shaped by lattice dispersion. We study leviton generation in finite one-dimensional tight-binding chains of non-interacting fermions driven by time-dependent voltage pulses. Using the single-particle density matrix, we resolve the excitation above the initial Fermi sea and quantify its quality through the average excitation number and its fluctuations. We find that clean leviton-like states emerge only in an intermediate regime where the pulse is slow enough to be resolved by the dynamics on the lattice, but not so slow that truncation and finite-size effects distort the Lorentzian profile in the time domain. Our time-resolved analysis shows that leviton formation is a coherent multi-mode process in which the first half of the pulse moves a unit charge into the excited subspace, creating substantial correlations with the Fermi sea, while the second half has the effect of suppressing these correlations and redistributing population among the low-energy states in the excited space, leading to a clean single-particle excitation exponentially localized in energy. Lorentzian pulses (even if truncated) systematically outperform non-Lorentzian pulse shapes in approaching the low-noise limit with increasing system size. We further identify finite-lattice signatures associated with band filling, pulse amplitude, linear voltage-drop geometry, and residual deviations from the continuum integer-charge condition. These results establish a microscopic framework for understanding leviton formation beyond the ideal continuum limit and for evaluating lattice-based platforms for coherent few-electron transport.
\end{abstract}

\maketitle

\section{Introduction}
The ability to generate on-demand coherent single-electron wave packets is a central goal of electron quantum optics (EQO) and an important step toward solid-state flying qubits \cite{GlRo16, Baeuerle18}. Many established single-electron sources rely on dynamically confining charge in mesoscopic structures, such as quantum dots, pumps, or turnstiles, where the emitted state is shaped by loading and ejection dynamics. An alternative route is to inject charge directly into a conductor by applying a voltage pulse to a contact. This avoids explicit charge capture in a localized potential, but a generic voltage drive perturbs the entire Fermi sea and creates additional neutral electron-hole pairs. Levitov and co-workers showed that this excess excitation can be eliminated by using Lorentzian voltage pulses injecting an integer number of electronic charges~\cite{Levitov1996, Ivanov1997}. The resulting minimal single-electron excitations, known as levitons, provide a clean and electrically controlled platform for electron quantum optics and solid-state flying-qubit architectures.

The experimental realization of levitons established voltage-pulse injection as a practical route to clean single-electron excitations. Lorentzian voltage pulses applied to a two-dimensional electron gas were shown to minimize partition noise at integer charge injection, confirming the suppression of excess electron-hole pairs predicted for minimal-excitation states~\cite{Dubois2013}. Subsequent tomography and two-particle interference experiments reconstructed the coherence of voltage-pulse-generated excitations and demonstrated their suitability for EQO, including Hong-Ou-Mandel interferometry~\cite{Jullien2014, Bocquillon2013, Bocquillon2014, Bisognin2019, GlRo16}. More recently, coherent control of leviton flying qubits in graphene quantum Hall edge channels has shown that these excitations can be manipulated as propagating electronic quantum states~\cite{Edlbauer2022, Assouline2023}.

Most theoretical descriptions of levitons are based on continuum scattering theory, chiral-edge models, or driven mesoscopic conductors, where the electronic spectrum is often linearized near the Fermi energy and the reservoirs are treated as macroscopic~\cite{KKL06, Bocquillon2014, GlRo16, Hofer2014}. These approaches successfully describe charge quantization, excess noise, coherence, and interference in conventional EQO geometries. However, growing interest in coherent single-electron dynamics and mesoscopic quantum transport has motivated the study of finite, discrete systems, including quantum-dot (QD) arrays \cite{Noiri2022}, graphene nanoribbons (GNRs) \cite{Wang2021}, and moiré superlattices \cite{Rickhaus2018}, as platforms for coherent quantum transport over mesoscopic length scales.

GNRs are particularly promising because their atomically precise width, edge termination, band structure, and localized spin degrees of freedom provide microscopic control over one-dimensional (1D) electronic transport~\cite{Cai2010, Ruffieux2016, Oteyza2022}. This tunability has motivated proposals for electron-optical elements based on GNRs, including crossed-ribbon beam splitters and mirrors, spin-polarizing beam splitters, and Mach–Zehnder-like interferometers formed from ribbon networks~\cite{Lima2016, Brandimarte2017, sanz2020crossed, Sanz2023, Sanz2024}. These developments suggest that levitonic wave packets could be shaped, split, and interfered within atomically engineered electronic circuits.

Finite lattice systems introduce effects that are absent or secondary in continuum descriptions, including nonlinear dispersion, finite bandwidth, filling-dependent propagation velocities, boundary reflections, and geometry-dependent scattering. Although tight-binding (TB) simulations have been used to study leviton-like excitations and their partitioning at quantum point contacts~\cite{Thomas2015}, a systematic understanding of how finite size, lattice dispersion, and band filling modify leviton generation remains underdeveloped.

In this work, we investigate how levitonic excitations are generated and characterized in finite 1D TB chains. We model a non-interacting fermionic lattice driven by time-dependent voltage pulses and follow the dynamics using the single-particle density matrix (SPDM), allowing us to resolve electron and hole excitations relative to the Fermi sea. We use the average excitation number $N_e$ and its fluctuation $\Delta N_e$ as primary figures of merit, supplemented by spectral, spatial, entropic, and current-based diagnostics of the excited-state density matrix. This framework is used to determine how finite chain length, pulse width, pulse amplitude, band filling, voltage drop geometry, and pulse shape affect the formation of leviton-like states. We find that clean single-electron excitations emerge only in an intermediate regime where the Lorentzian pulse is sufficiently resolved by the time scale of dynamics on the lattice while remaining compatible with the finite pulse duration.

The remainder of the paper is organized as follows. Section \ref{sec:Model} introduces the finite TB model, the truncated voltage pulses used to generate excitations, and the SPDM tools used to quantify leviton formation. Section \ref{sec:results} presents the numerical results, focusing on the dependence of excitation quality on pulse parameters, system size, filling, pulse shape, and linear voltage-drop geometry. Section \ref{sec:discussion} discusses the implications of these finite-lattice effects for experimentally relevant platforms, and Section \ref{sec:conclusions} summarizes the main conclusions.

\section{Model}
\label{sec:Model}

\subsection{System description}

\begin{figure}
    \centering
    \includegraphics[width=0.45\textwidth]{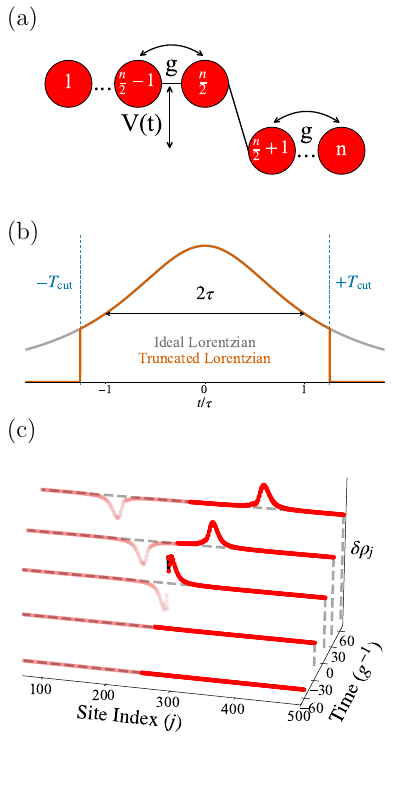}
    \caption{
    Leviton formation on a TB chain. 
    {(a)} Sketch of the setup, defining the TB model and the voltage profile along the chain.
    {(b)} The time-dependent voltage pulse is applied to the left half-chain (top), with a Lorentzian profile characterized by the half-width at half maximum $\tau$ (gray). For finite pulses we define a pulse cutoff, $T_\text{cut}$ (red).
    {(c)} Snapshots of leviton (opaque) and anti-leviton (transparent) generation for a $n=500$ site chain as shown by the deviation in charge density.
    }
    \label{fig:setup}
\end{figure}
 
The system, sketched in \Figref{fig:setup}(a), is described by the Hamiltonian
\begin{equation}
\mathscr{H}(t) = \mathscr{H}_0 + \mathscr{H}_{1}(t),    \label{eq:fullH}
\end{equation} 
where $\mathscr{H}_0$ represents the static lattice and $\mathscr{H}_1(t)$ accounts for the time-dependent bias at time $t$.
The static system is modeled by a nearest-neighbor, spinless, TB Hamiltonian,
\begin{equation}\label{eq:H0}
    \mathscr{H}_0 = - g\sum_{\langle j,l \rangle}  (c_j^\dagger c_l + c_l^\dagger c_j),
\end{equation}
where $c_j^\dagger$ ($c_j$) is the creation (annihilation) operator for an electron at site $j$, where the on-site energy is set to zero, and $g$ is the hopping amplitude between nearest-neighbor sites. Throughout this work, we consider a homogeneous chain with $n$ sites, uniform hopping $g$, and $\hbar=1$.

We focus on finite linear chains with open boundary conditions, motivated by quasi-1D platforms such as quantum-dot arrays and GNRs. In this geometry, the voltage can be applied to a segment that terminates at one boundary, leaving a single voltage-drop region in the system and producing a predominantly unidirectional excitation. In a closed chain, by contrast, any finite biased segment introduces two voltage-drop regions, one at each end, and therefore generically excites both left- and right-moving wave packets. Open boundaries thus provide the simplest setting for isolating the formation and propagation of a single leviton-like excitation.  

Electronic excitations are generated by applying a time-dependent voltage $V(t)$ to a contiguous subset of sites $\mathcal{L}$, which, in this work, is always in the ``left" half of the chain, $j \in \{1, \dots, n/2\}$ as shown in \Figref{fig:setup}(a). The voltage drop is thus chosen to occur in the middle of the chain. The corresponding interaction Hamiltonian for a voltage drop across a single bond is 
\begin{equation}\label{eq:H1}
\mathscr{H}_1(t) =  \sum_{j \in \mathcal{L}} V_j(t)c_j^\dagger c_j.
\end{equation}

In the quasistationary continuum theory, the current responds proportionally to the applied voltage, $~{I(t) = e^2/h\, V(t)}$, where $e$ is the fundamental electric charge and $h$ Planck's constant. Minimal particle-hole excitation states are generated by the area-quantization condition   
$\int V dt =   n_L h/e$,
where $n_L$ is an integer \cite{Levitov1996, KKL06, Dubois2013}. For $n_L=1$ the pulse ideally creates a single-electron excitation above the Fermi sea without additional electron-hole pairs.

An ideal Lorentzian pulse has long (infinite) temporal tails and must therefore be truncated in practical implementations, yielding a finite-duration pulse suitable for leviton generation. The boundary conditions of the finite system provide additional motivation if one wishes to avoid reflections at the chain edges of the traveling leviton. We therefore introduce the truncated Lorentzian (LT) pulse with a cutoff time $T_\text{cut}$ for the applied pulse, as shown in \Figref{fig:setup}(b): 
\begin{equation}\label{eq:LT}
V^\text{LT}(t)=
n_L \dfrac{2\tau}{t^2+\tau^2} \theta (T_{\rm cut} - |t|),
\end{equation}
with $\theta(x)=1$ for $x\geq0$ and $0$ otherwise.
We choose $gT_\text{cut}=n/8$ to ensure the excitation propagates midway along the half-chain by the time the pulse is completed. This choice is motivated by a voltage drop at the center of the chain and a leviton group velocity around $2g$ (in units of the lattice spacing). 
Such a characteristic situation is illustrated in \Figref{fig:setup}(c).
For comparison, we also perform calculations using other voltage pulses described in \Appref{app:pulseshape}.

In the finite lattice studied here, $n_L$ is treated as a tunable parameter that characterizes the pulse amplitude. Unless otherwise stated, we use the conventional single-leviton pulse ($n_L=1$). A direct comparison between Lorentzian, Gaussian, and square pulses is presented in \Secref{sec:results}.

\subsection{Computation of unitaries and time-propagated states}

The numerical results presented below are obtained by evolving the SPDM under the time-dependent Hamiltonian in \Eqref{eq:fullH}. Unless stated otherwise, the system is initialized in the half-filled ground state of $\mathscr{H}_0$
corresponding to a zero-temperature Fermi sea with filling fraction $\nu=0.5$.
The pulse is applied over the time interval $[-T_\text{cut}, T_\text{cut}]$, after which the Hamiltonian becomes time independent and the excitation propagates freely through the lattice.  Unless shown as an explicit function of time, they are evaluated at $t = T_\text{cut}$, when the pulse has completely vanished and the excited state is fully formed. Since the derived quantities $N_e$ and $\Delta N_e$ are conserved under $H_0$ they characterize the excitation for all $t>T_\text{cut}$. 

The dynamics generated by \Eqref{eq:fullH} is described by the time-ordered unitary
\begin{equation}
U(t,t_{i})=\mathcal{T}\exp\left[ -i \int_{t_i}^{t}H(s) \;ds\right],
\label{eq:unitary}
\end{equation}
where $\mathcal{T}$ is the time-ordering operator and $H(s)$ is the single-particle Hamiltonian corresponding to \Eqref{eq:fullH}.
We evaluate it by discretizing time into uniform steps of size $\Delta t$ and approximating the propagator as a product $U(t,t_i)\approx \prod_\ell U(t_{\ell + 1}, t_\ell)$.

As detailed in \Appref{app:numerics}, in each interval we retain the first- and second-order Magnus terms, which describes propagation under an effective, time-independent Hamiltonian $H_{\text{eff}, \ell}$ that approximates the exact propagator to fourth order in $\Delta t$. To avoid repeated matrix exponentiation, we employ a rational (Padé-type) approximation which uses $e^{-iA}=e^{-iA/2}\left(e^{+iA/2}\right)^{-1}$ and then approximates both exponentials by the first few terms of the power series. We use 
\begin{equation}
    e^{-i\Delta t H_{\text{eff},t_{\ell}}/2} \approx 1+B_4 \equiv 1+\sum_{m=1}^4\frac{1}{m!} \left[-\frac{i\Delta t}{2}H_{\text{eff},t_{\ell}}\right]^m,
\end{equation}
and thus 
\begin{equation}
    e^{-i\Delta t H_{\text{eff},t_{\ell}}} \approx (1+B_4)(1+B_4^\dagger)^{-1},
\end{equation}
which preserves unitarity and maintains the same order of accuracy.

We also use a fixed time step $\Delta t$ chosen sufficiently small to ensure convergence of all observables. A convenient upper bound for $\Delta t$ is obtained by requiring that the discretization resolves all relevant time scales of the problem, namely the pulse duration $T_\text{cut}$, pulse width $\tau$, voltage amplitude $V_\text{max}$, and hopping strength $g$. This leads to the estimate $\Delta t \lesssim \min \left(\frac{T_\text{cut}}{10}, \frac{\tau}{10}, \frac{1}{10\;V_\text{max}}, \frac{1}{10\,g}\right)$, which is used as a guideline to select a sufficiently small time step. Convergence with respect to the time step and system size was verified, and the results reported here are insensitive to further refinement of the numerical parameters.

\subsection{Theoretical tools to characterize leviton states}
Due to the quadratic form of the Hamiltonian  in \Eqref{eq:fullH} with respect to the fermionic operators, the many-body dynamics can be expressed entirely in terms of the SPDM
\begin{equation}
    \varrho_{k\ell} = \langle c_k^\dagger c_l\rangle_\rho,
\end{equation}
where $\rho$ is the full many-body density matrix.
Its time evolution follows $\varrho(t)=U(t, 0)\, \varrho(0) \, U^\dagger(t, 0)$. The SPDM are normalized such that their trace is equal to the expected particle number, $\tr{\varrho}=N$.

To characterize the injected excitation, we partition the single-particle Hilbert space into the initially occupied and empty subspaces of $H_0$. Let $P_{\mathrm{occ}}$ denote the projector onto the occupied eigenstates of $H_0$ and $P_\mathrm{emp}=I-P_{\mathrm{occ}}$ the complementary projector onto the empty subspace. 
The system is initialized at $t=t_0$ (which we choose as $t_0=-T_\text{cut}$) in the zero-temperature Fermi sea corresponding to a filling fraction $\nu = \tr{P_\text{occ}}/n$, such that
\begin{equation}
    \varrho(t_0)=P_\mathrm{occ}.
\end{equation}
We then define the excited-space SPDM (e-SPDM) as
\begin{equation}\label{eq:Xdensitymatrix}
    \varrho_e(t)=P_{\mathrm{emp}}\, \varrho(t) \, P_{\mathrm{emp}} .
\end{equation}
Since the initial state is given by the projector $P_{\mathrm{occ}}$, the e-SPDM can equivalently be written as 
\begin{equation}
    \varrho_e(t) = P_{\mathrm{emp}}\,U(t,t_0) P_{\mathrm{occ}}U(t,t_0)^\dagger\, P_\mathrm{emp}.
\end{equation}
This representation emphasizes that the voltage pulse effectively rotates the initially occupied subspace within the single-particle Hilbert space, generating particle and hole excitations relative to the original Fermi sea.

Figure~\ref{fig:setup}(c) shows the spatial profile of the leviton (opaque) and anti-leviton (transparent) in a chain consisting of $n=500$ sites. The profile is determined from the difference in population between the time-evolved density matrix $\rho(t)$ and the initial density matrix $\rho_0$ at each site, \ie, $\delta\rho_j=\rho_j(t)-\rho_{0,j}$. The applied pulse has a width characterized by $g\tau = 4.0$ and a cutoff time $gT_\text{cut}=62.5$ corresponding to a length of $n/8$--avoiding any interference from reflections at the chain boundaries and placing the peak of the excited charge distribution in the center of the right half of the chain.

To quantify the generated excitation, we introduce two figures of merit. The expected number of particles promoted above the Fermi sea is
\begin{equation}
    N_e(t) = \tr{\varrho_e(t)} = \tr{P_{\mathrm{emp}}\,\varrho(t)},
\end{equation}
while the standard deviation
\begin{equation}\label{eqn:varNe}
    \Delta N_e = \sqrt{\tr{\varrho_e}-\tr{\varrho_e^2}},
\end{equation}
measures the deviation from an excitation with sharp integer particle number (see a derivation in \Appref{app:varNe}).

For an ideal single leviton, the e-SPDM is a rank-one projector, yielding $N_e=1$ and $\Delta N_e = 0$. In numerical simulations, we instead observe one dominant eigenvalue of $\varrho_e$ close to unity accompanied by smaller contributions from additional modes, reflecting residual electron-hole excitations. $\Delta N_e$ will be our primary figure of merit for the generated excitation.

\section{Results}
\label{sec:results}

In this section we analyze how the Lorentzian pulse parameters ($\tau$ and $n_L$) and the chain length ($n$) influence the excitation dynamics through their impact on $N_e$ and $\Delta N_e$.  We focus on moderate chain lengths ($n$ on the order of a few 100 to a few 1000) where the levitonic properties are already well-expressed, but finite-size effects still conspicuous. For smaller $n$, Lorentzian pulses lead to poor single-particle excitations (both in terms of purity and in terms of localization) and for larger $n$ the excitations approach (slowly, but as expected) the continuum limit behavior.

\begin{figure}
    \centering
    \includegraphics[width=0.5\textwidth]{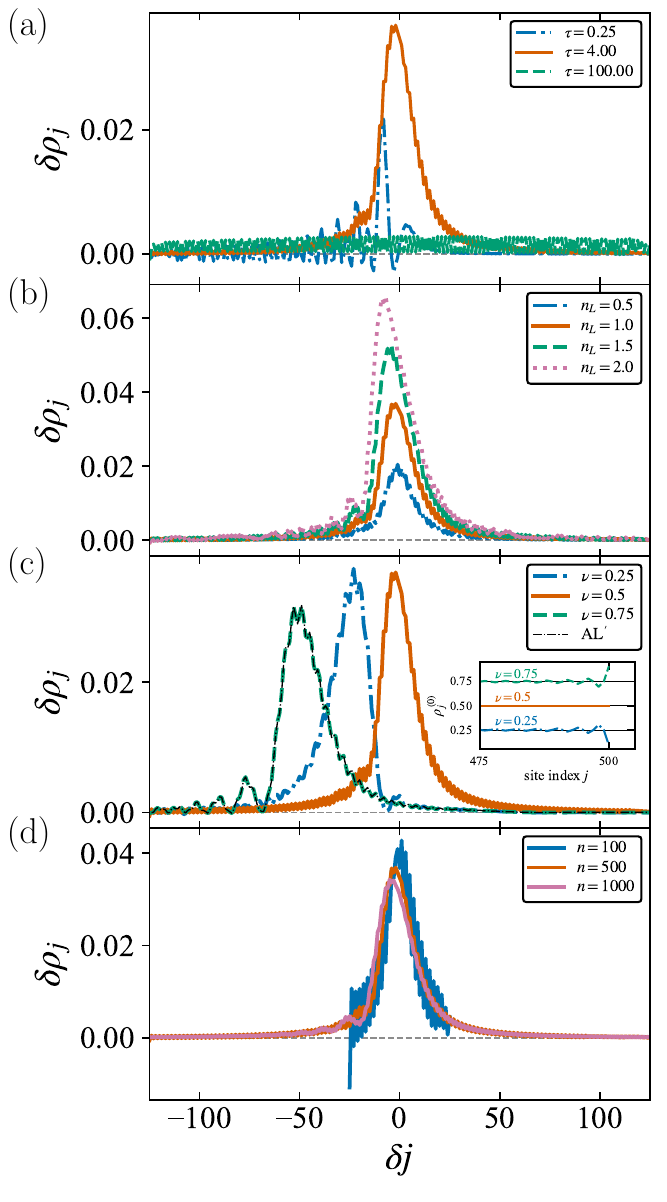}
    \caption{
    Induced charge-density difference after a completed Lorentzian pulse as described by \Eqref{eq:LT}.
    {(a)} $g\tau=0.25$ (blue, dot-dash), $g\tau=4$ (orange, solid), and $g\tau=10$ (green, dash) show that the quality of the charge density profile depends strongly on the chosen width of the pulse, with $g\tau=4$ resulting in a reasonable Lorentzian profile.
    {(b)} The pulse amplitude scale factors $n_L=0.5$ (blue, dot-dash), $n_L=1$ (orange, solid), $n_L = 1.5$ (green, dash), and $n_L=2$ primarily impact the amplitude of the charge density profile while also introducing asymmetry.
    {(c)} Filling factor $\nu=0.25$ (blue, dot-dash), $\nu=0.5$ (orange, solid), and $\nu=0.75$ (green, dash) lead to different Fermi velocities. Initial charge density is uniform only for $\nu=0.5$ (inset). The $\nu=0.75$ leviton profile matches the $\nu=0.25$ antileviton (AL) profile (black, dot-dash).
    {(d)} Charge density profile as chain length is varied. The amplitude of the profile decreases as the chain length increases.
    Unless otherwise stated, the parameters used are $n=500$, $\nu=0.5$, $g\tau=4$, $gT_\text{cut}=n/8$, $w=1$, and $n_L=1$.
    }
    \label{fig:ChargeDensity}
\end{figure}

\subsection{Generated charge density profile} 
Figure \ref{fig:ChargeDensity} shows the difference in charge density profile, $\delta\rho_j$, with panel (a) focusing on an LT pulse for an $n=500$ site chain at half filling over a representative range of pulse widths $\tau$. The difference in charge density is striking. For $g\tau=4$, the Lorentzian pulse has a total width of roughly 16 sites at half the maximum amplitude (FWHM), which occupies approximately 6\% of the entire half-chain. As a result, the charge density presents a Lorentzian shaped profile. If the half-width of the pulse is reduced to $g\tau=0.25$, the FWHM is reduced to a single site. A pulse this narrow cannot be resolved by the discrete chain and the resulting charge density is distinctly non-Lorentzian. Conversely, if the pulse half-width is $g\tau=100$, the FWHM is roughly 400 sites, which is 1.6 times larger than the half-chain itself. The charge density profile then resembles a broad, low-amplitude Lorentzian according to \Eqref{eq:LT} with the addition of strong oscillations. The drastic difference in charge density profile seen in \Figref{fig:ChargeDensity}(a) suggests that in finite systems, an optimal value exists for the characteristic pulse width $\tau$.

Figure~\ref{fig:ChargeDensity}(b) shows the trend in charge density with respect to the Lorentzian pulse amplitude $n_L$. As one would expect, increasing the pulse amplitude directly leads to an increase in the charge density amplitude.
Figure~\ref{fig:ChargeDensity}(c) shows the charge-density profiles generated at different fillings. 
For the nearest-neighbor TB chain, the Fermi velocity is maximal at half filling, $v_\mathrm{F}^{(1/2)}=2ga$, where $a$ is the lattice constant. 
At quarter and three-quarter filling it is reduced to $v_\mathrm{F}^{(1/4)}=v_\mathrm{F}^{(3/4)}=\sqrt{2}ga$. 
One therefore expects the half-filled excitation to propagate furthest along the chain at a fixed observation time, as seen in \Figref{fig:ChargeDensity}(c). 
However, the maxima of the $\nu=0.25$ and $\nu=0.75$ profiles do not coincide, despite their equal Fermi velocities. 
This reflects the fact that the leviton is a finite-bandwidth wave packet formed from a superposition of single-particle eigenstates, rather than a monochromatic excitation moving exactly at $v_\mathrm{F}$. 
For states just above the Fermi level, the group velocities are larger than $\sqrt{2}ga$ at $\nu=0.25$, but smaller than $\sqrt{2}ga$ at $\nu=0.75$. 
Consequently, the $\nu=0.75$ density maximum lags behind the corresponding $\nu=0.25$ maximum.

The shape of the density profile is also controlled by the local curvature of the band near the Fermi level. 
At half filling, the Fermi point coincides with the inflection point of the TB dispersion, so the group velocity of particle-like states above the Fermi level varies only quadratically with their wave number measured from $k_\mathrm{F}$. The finite-bandwidth leviton therefore samples a comparatively uniform velocity distribution, which helps preserve a nearly symmetric Lorentzian charge-density profile. 
Away from half filling, this protection is lost: the group velocity varies linearly with wave number near $k_\mathrm{F}$, producing stronger velocity shear across the spectral components of the wave packet and hence an asymmetric charge-density profile. The filling factor also influences the initial charge density. For half filling, the initial charge density is homogeneous, but when the filling deviates from $\nu=0.5$, Friedel oscillations are visible near the chain boundaries (inset).

At half filling, particle--hole symmetry relates the leviton and antileviton density profiles by spatial reflection. 
This symmetry is broken away from half filling, as illustrated by the thin black dot-dashed curve in \Figref{fig:ChargeDensity}(c), which shows that the reflected $\nu=0.25$ antileviton profile coincides with the $\nu=0.75$ leviton profile.

Finally, \Figref{fig:ChargeDensity}(d) is a comparison of a fixed-width LT pulse ($g\tau=4$) and a cutoff that scales with chain length $gT_\text{cut}=n/8$. As the chain length increases, less of the Lorentzian voltage pulse is truncated and the resulting excitation becomes more Lorentzian. It should be noted, however, that the excited charge density remains clearly asymmetric and we see no indication of convergence to a Lorentzian. This is as expected since only for $g\tau\gg1$ when the scattering time is short compared with the pulse duration, as assumed in the continuum case \cite{Levitov1996}.

\subsection{Characterization of the time evolution of the excited subspace density matrix}

\begin{figure*}
    \centering
    \includegraphics[width=0.8\textwidth]{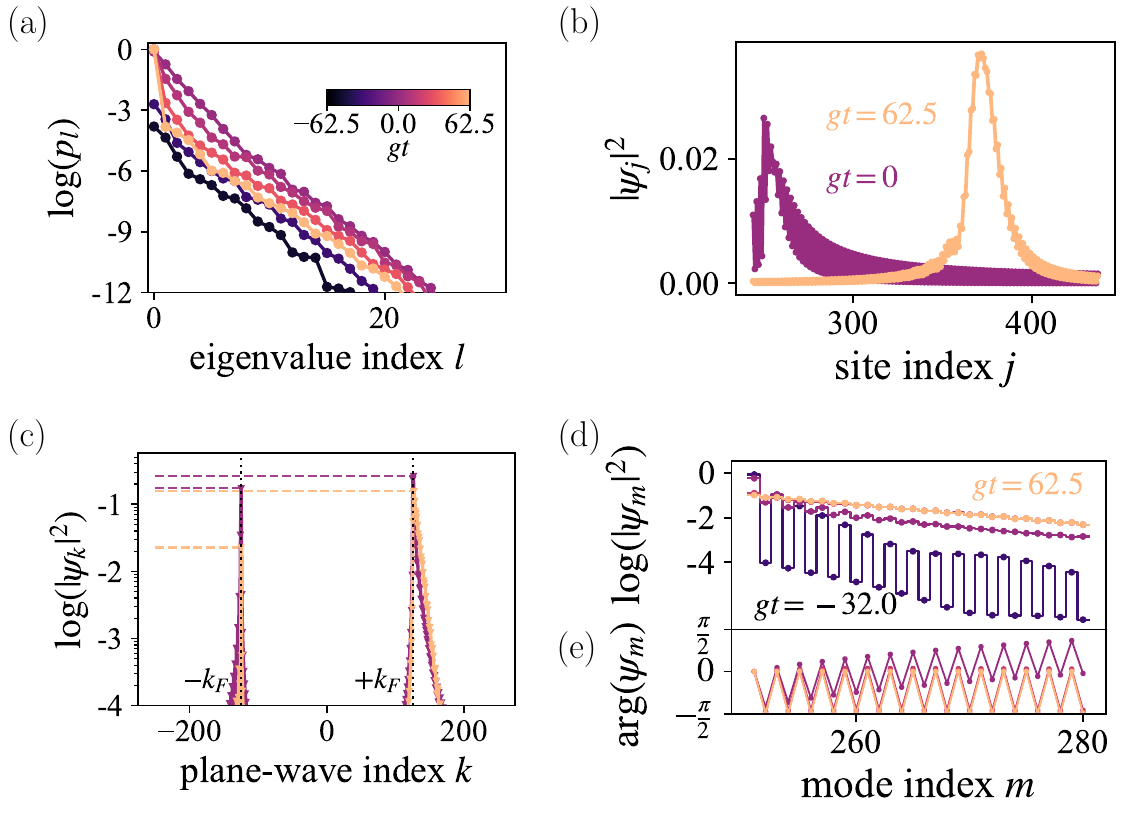}
    \caption{
    Excited-space density matrix $\varrho_e$ and leviton wave function $\psi$ during pulse-driven excitation in a $n=500$ TB chain. All panels refer to a Lorentzian pulse with $g\tau=4$, $gT_{\rm cut}=62.5$, $n_L=1$, and voltage-drop width $w=1$. Colors indicate the dimensionless time $gt$.
    (a) Eigenvalue spectrum $p_l$ of $\varrho_e$ at $gt=-50,-32,0,10,31,62.5$, with corresponding $(N_e,\Delta N_e)=(0.0002,0.0144)$, $(0.0024,0.0487)$, $(0.9960,0.6098)$, $(1.0406,0.2038)$, $(1.0025,0.0541)$, and $(1.0001,0.0221)$, respectively. The spectrum evolves toward a nearly rank-one excitation after the pulse.
    (b) Site probability $|\psi_j|^2$ of the leviton wave function at $gt=0$ and $gt=62.5$, with $N_e=0.9960$ and $N_e=1.0001$, showing propagation from the voltage-drop region into the right half of the chain.
    (c) Plane-wave probability $|\psi_k|^2$ at the same two times; dotted vertical lines mark $\pm k_F$. The final wave function is concentrated near and above $+k_F$.
    (d) Energy-mode probability $|\psi_m|^2$ of the leviton wave function in the eigenbasis of $H_0$ at $gt=-32,0,15,62.5$, together with (e) the corresponding residual phase $\arg(\psi_m)$ after removing the free dynamical phase at $gt=0,15,31,62.5$. The final-time energy distribution has an approximately exponential envelope, while the phase shows an even-odd pattern.}
    \label{fig:wavefunction}
\end{figure*}

The charge-density profiles in \Figref{fig:ChargeDensity} provide a spatially intuitive picture of the generated excitations, but they do not by themselves determine whether the excitation is a clean leviton. A density maximum can arise  even when the excited subspace contains several partially occupied modes, corresponding to additional particle-hole excitations. To resolve this structure, we analyze the eigenvalues and eigenvectors of the excited-space density matrix $\varrho_e(t)$ defined in \Eqref{eq:Xdensitymatrix}. For an ideal single leviton, $\varrho_e$ is a rank-one projector with a single eigenvalue equal to unity and all others zero. In a finite lattice, deviations from this structure provide a direct measure of the fragmentation of the excitation into additional modes. 

Figure~\ref{fig:wavefunction} illustrates this process for a representative LT pulse with $n=500$, $g\tau=4$, $gT_\text{cut}=62.5$, $n_L=1$, and $w=1$. Figure~\ref{fig:wavefunction}(a) shows the spectrum of $\varrho_e(t)$ at several times during the pulse. Initially, the excited subspace is essentially empty. During the first half of the pulse, population is transferred into the initially empty sector and several eigenvalues grow. At the center of the pulse, the total number of excited particles has overshot unity and the impurity is large, indicating that the state is not yet a clean single-mode excitation. During the second half of the Lorentzian pulse, however, the subleading eigenvalues are strongly suppressed while the largest eigenvalue remains close to one. Thus, leviton formation is not simply the monotonic injection of charge into the empty subspace; it is a coherent process in which the second half of the pulse removes much of the unwanted particle-hole content generated earlier in the drive.

We identify the eigenvector corresponding to the largest eigenvalue of $\varrho_e$ as the instantaneous leviton wave function $\psi$. Figure~\ref{fig:wavefunction}(b-c) show this wave function in the site and plane-wave bases at the center of the pulse (purple) and after the pulse has ended (light orange). At $gt=0$, the wave function is localized near the voltage drop and contains appreciable weight near both Fermi points. By $gt=62.5$, it has propagated into the right half of the chain and forms a Lorentzian-like wave packet whose plane-wave distribution is concentrated near and above $+k_F$. This redistribution in plane-wave space confirms that the final state is predominantly right-moving, despite the presence of both positive and negative plane-wave components during the formation stage associated with the standing wave eigenstates of the finite chain.

Figure~\ref{fig:wavefunction}(d-e) gives a complementary view in the eigenbasis of the static Hamiltonian $H_0$. Figure~\ref{fig:wavefunction}(d) shows that, after the pulse, the leviton wave function is distributed over many modes above the Fermi level with an approximately exponential envelope, as expected for a Lorentzian excitation. Figure~\ref{fig:wavefunction}(e) shows the corresponding phase after removing the free dynamical phase. Both modulus and phase of the amplitudes change markedly during the first half of the pulse, until the lowest unoccupied mode acquires population close to 1. At $t=0$ there is still a clear over-population of the lowest unoccupied level and "missing" amplitudes at higher energies, but after one quarter of the second half of the full truncated pulse the amplitudes are already almost indinstinguishable from the final state, see the almost fully overlapping curves for $gt=15$ vs. $gt=62.5$ in \Figref{fig:wavefunction}(d). In particular, the phase pattern becomes nearly time-independent after removal of the free dynamical phase, indicating that the internal energy-mode structure of the wave packet is largely fixed.

\subsection{Characterization of generated leviton in the excited subspace}
\begin{figure}
    \centering
    \includegraphics[width=0.5\textwidth]{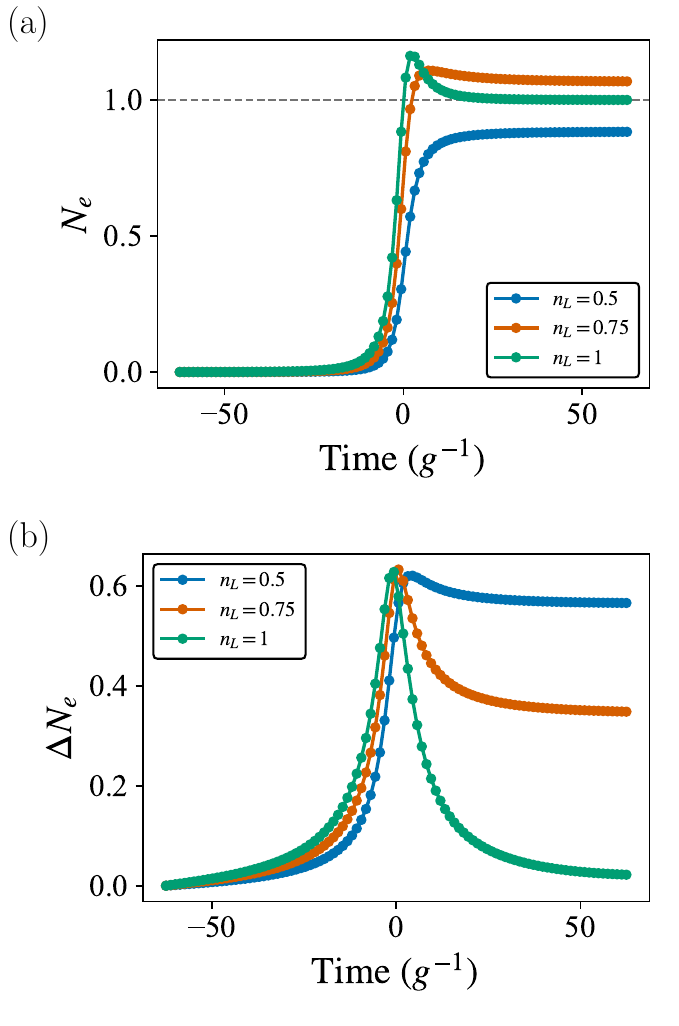}
    \caption{
    {(a)} Average number of excitations $N_e$ and
    {(b)} standard deviation $\Delta N_e$ as a function of time for a Lorentzian pulse with $n_L=1$ (green), $n_L=0.5$ (blue), and $n_L=0.75$ (orange). For integer pulse charge $n_L=1$, the number of excitations approaches an integer value and the standard deviation is reduced.
    The parameters used are $n=500$, $g\tau = 4$, $gT_\text{cut}=62.5$, $\nu = 0.5$, and $w=1$.}
    \label{fig:NeDelNeVStime}
\end{figure}

Figure \ref{fig:NeDelNeVStime} shows the time evolution of $N_e$ and $\Delta N_e$ for an $n=500$ chain at half filling. The pulse width is characterized by $g\tau=4$ and $gT_\text{cut}=n/8=62.5$. In each case, the system begins in the ground state with $N_e=0$. Once the pulse amplitude becomes comparable to the discretization gap, population transfer takes place. Since the quantity $N_e$ measures the total occupation transferred into the initially empty subspace of $H_0$, not the net injected charge during the pulse, $N_e(t)$ may overshoot due to coherent mixing between the occupied and empty eigenstates of the static Hamiltonian. This can be seen in \Figref{fig:NeDelNeVStime}(a) for $n_L=0.75$ (red) and $n_L=1$ (green). Moreover, when the pulse area deviates from the integer leviton condition, additional electron-hole pairs are generated, so that the final value of $N_e$ can exceed unity even for $n_L < 1$, as can be seen for $n_L=0.75$. Figure \ref{fig:NeDelNeVStime}(b) corroborates the generation of additional electron-hole pairs for non-integer $n_L$ through an increase in the standard deviation $\Delta N_e$, reaching a maximum near the mid-point of the pulse. The area quantization condition $~{\int V\;dt = n_Lh/e}$ manifests in the finite system as a decrease in the standard deviation of $N_e$ as $n_L$ moves towards an integer value. For the infinite chain, the $n_L=1$ pulse is expected to generate a pure levitonic state with $N_e=1$ and $\Delta N_e=0$ after the pulse has completed. It is important to note that the standard deviation at the final time is still non-zero for $n_L=1$ for the $n=500$ chain, which likely stems from the deviation of the LT pulse from an ideal Lorentzian.

\subsection{Role of Lorentzian pulse parameters}

\begin{figure*}
    \centering
    \includegraphics[width=0.8\textwidth]{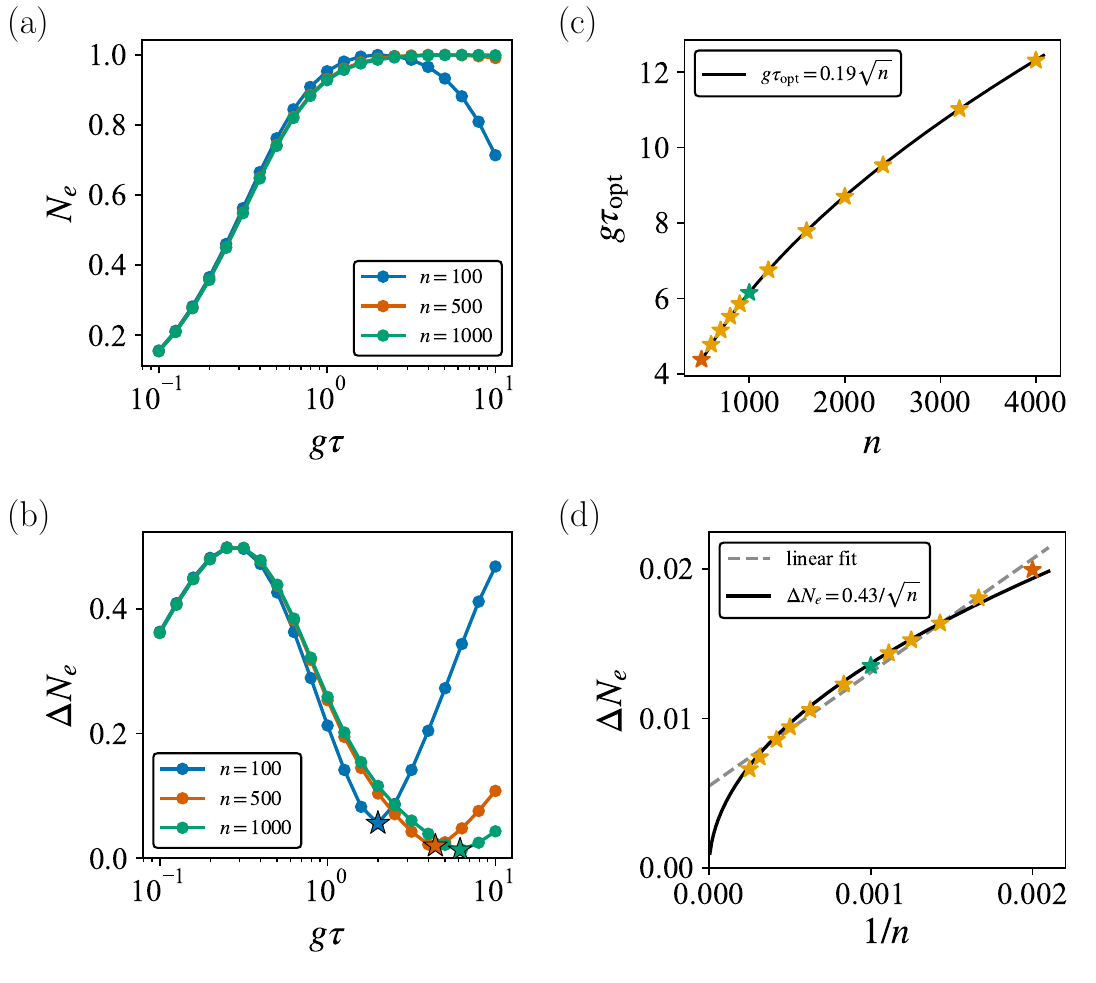}
    \caption{
    {(a)} Average number of excitations $N_e$, and 
    {(b)} standard deviation $\Delta N_e$ as a function of $g\tau$ for Lorentzian pulses in chains of length $n=100$ (blue), $n=500$ (orange), and $n=1000$ (green). Performance at small $g\tau$ is limited by the sudden-drive regime, while at large $g\tau$ it is limited by pulse truncation. In (b), The extracted minima at $\tau=\tau_\text{opt}$ are indicated by stars.
    {(c)} Extracted optimal width $\tau_\mathrm{opt}$ as a function of chain length. The trend is well described by $g\tau_\mathrm{opt}=0.19\sqrt{n}$. 
    {(d)} Extracted minima $\Delta N_e(\tau_\mathrm{opt})$ as a function of inverse chain length $1/n$, together with a linear (gray dashed) and $1/\sqrt{n}$ (black) scaling.
    The parameters used are $gT_\text{cut}=n/8$, $\nu=0.5$, and $w=1$.}
    \label{fig:NeDelNeVStau_chainlength}
\end{figure*}

It is not the aim of this work to comprehensively explore the pulse-parameter space, but to highlight some finite-size-induced trends and deviations form the ideal case and to identify favorable parameter combinations that allow to generate excitations with especially leviton-like properties. As we shall see, typically low fluctuation in particle number (around an integer $>0$) guides the way to ensure also the expected $\tau$-related peaked distributions in position and plane-wave space. 

In \Figref{fig:ChargeDensity}(a), the dramatic dependence of the charge density profile on choice of pulse width $\tau$ is an indication that, for finite systems, there exists an optimal width, $\tau_\text{opt}$. In the following, our main focus will be on the effects of $\tau$ and system size. To facilitate the comparison across different system sizes, we fix $gT_\text{cut}=n/8$ for most of the discussion below. 

Figure~\ref{fig:NeDelNeVStau_chainlength}(a-b) shows the average number of excitations generated and the standard deviation as a function of the pulse width, respectively. At $\tau=\tau_\text{opt}$, the LT pulse generates $N_e\approx1$ with a minimized $\Delta N_e$. 

The pulses with widths $\tau\lesssim\tau_\text{opt}$ under perform due to a competition between the excitation generated by the pulse and the speed at which the excitations can propagate away from the source, set by $H_0$. At half-filling, excitations travel along the chain at a speed approximately equal to the magnitude of the Fermi velocity, $\lvert v_\text{F}\rvert=2ga$. The central width of the pulse is then spread over $4g\tau a$. For $g\tau=0.5$, the main width of the pulse is spread over two sites. This non-Lorentzian profile exhibits non-integer $N_e$ with a large standard deviation across all calculated chain lengths. 
In the extreme limit $\tau\ll\tau_\text{opt}$, the standard deviation begins to drop, and tends to zero as $\tau\to 0$. This regime represents a sudden approximation, where pulses do not impart changes on the state of the discrete chain. The corresponding unitary in this regime is equal to the identity for voltage drops across a single bond, as they impart a phase of $2\pi$ to the sites exposed to the voltage pulse and of 0 to the rest. Consequently, $N_e$ also tends to zero.

In the opposite regime, where $\tau\gtrsim\tau_\text{opt}$, we again notice a decrease in $N_e$ and a corresponding increase in $\Delta N_e$. In this regime, the excitations have ample time to propagate away from the source, but the fixed value of $T_\text{cut}$ limits the underlying voltage profile that is generated. This pushes the generated pulse profile further from a Lorentzian leading to an increase in $\Delta N_e$. 
The tension between the propagation speed of the excitations and the fidelity between the LT pulse and an ideal Lorentzian establish an optimal intermediate regime in which the pulse timescale is matched to the dynamical scales of the system, yielding a near-integer and minimally dispersed excitation.

Figure~\ref{fig:NeDelNeVStau_chainlength}(c) shows the extracted values for $\tau_\text{opt}$ as the system size increases from $n=500$ to $n=4000$. The trend is clearly not linear, with $\tau_\text{opt}$ increasing more quickly for smaller system sizes and slowing down as system size surpasses $n=1000$. The rate at which $\tau_\text{opt}$ increases with system size is surprising. The observation that $\tau_\text{opt}$ only increases by a factor of 2 as the system size increases by nearly an order of magnitude suggests that the tails of the Lorentzian profile play an important role once the central region of the profile has spread over an adequate number of sites.

Figure~\ref{fig:NeDelNeVStau_chainlength}(d) shows the corresponding standard deviation for simulations at $\tau=\tau_\text{opt}$ as a function of inverse system size. Longer chains are able to host more complete Lorentzian charge density profiles without introducing complications from the boundary. It is expected that as the system size increases, $\Delta N_e(\tau_\text{opt})$ will decrease. As the chain length tends to infinity, $\Delta N_e$ is expected to reach zero. However, our simulations are sufficiently far from this limit. Noticeably, the trend is non-linear and is well-approximated by a square-root  dependence on $n$.

\begin{figure}
    \centering
    \includegraphics[width=0.5\textwidth]{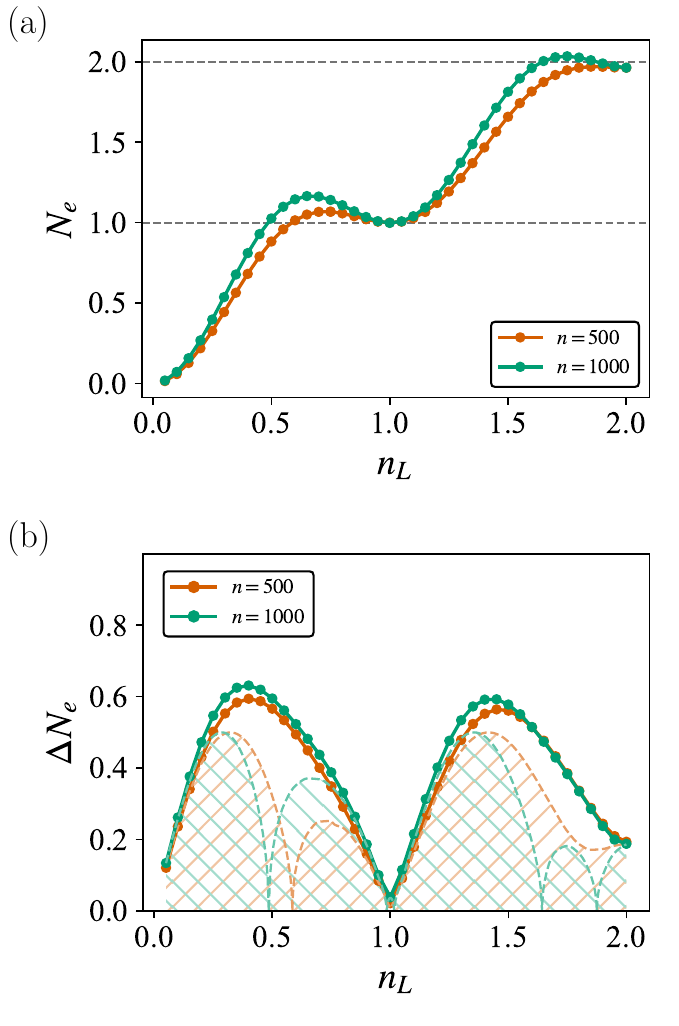}
    \caption{
    {(a)} Average number of excitations $N_e$ and
    {(b)} standard deviation $\Delta N_e$ as a function of $n_L$ for Lorentzian pulses in chains of length $n=500$ (orange, solid) and $n=1000$ (green, solid). In both cases, $\Delta N_e$ is minimized near integer values of $n_L$, in accordance with the ideal leviton expectation. The dashed curves in (b) show the corresponding lower bound set by the distance of $N_e$ to the nearest integer. The parameters used are $g\tau = 4$, $gT_\text{cut} =n/8$, $\nu = 0.5$, and $w=1$.}
    \label{fig:NeDelNeVSnL}
\end{figure}

Figure~\ref{fig:NeDelNeVSnL} summarizes the dependence of the excitation number $N_e$ and its standard deviation $\Delta N_e$ on the pulse parameter $n_L$ for fixed systems of size $n=500$ (orange) and $n=1000$ (green). As shown in \Figref{fig:NeDelNeVSnL}(a), $N_e$ increases overall with $n_L$, but exhibits clear non-monotonic behavior, particularly in the vicinity of integer $n_L$ values. In these regions, increasing the pulse amplitude does not lead to a strictly monotonic increase in the excited-state population; instead, $N_e$ can decrease slightly before continuing its overall growth. We observe an increase in the local maxima as system size increases. This behavior reflects the coherent nature of the excitation process, and the fact that non-integer pulses create a diverging amount of noise in large-$n$ systems. The final occupation of the excited manifold is determined by interference between multiple particle-hole excitation channels rather than by a simple additive injection of charge.  

The corresponding behavior of the standard deviation $\Delta N_e$, shown in \Figref{fig:NeDelNeVSnL}(b), further clarifies this picture. Clear minima in $\Delta N_e$ occur at integer values of $n_L$, consistent with the expectation that quantized pulses produce nearly pure single-particle excitations. We find a slight improvement in $\Delta N_e$ for $n=1000$, which is most likely due to the corresponding increase in $T_\text{cut}$, and agrees with the trends seen in \Figref{fig:NeDelNeVStau_chainlength}.

To quantify how close the generated excitation is to the minimal-fluctuation state compatible with $N_e$, we compare $\Delta N_e$ to the lower bound \footnote{For a state with $N_e=n_0+x$ with $n_0$ an integer and $x\in[-1/2,1/2]$, the standard deviation of $N_e$ is at least $\Delta_\mathrm{min}=\sqrt{|x|(1-|x|)}$. The bound is tight since the mixed state $(1-|x|)\ketbra{n_0}{n_0}+|x|\ketbra{n_0\pm1}{n_0\pm1}$ has particle number expectation value $n_0+x$ and variance $|x|(1-|x|)$. The sign in the projector in the second term is given by the sign of $x$.} $ \Delta_{\mathrm{min}}$ (dashed curve), which represents the minimal possible variance compatible with a given mean excitation number $N_e$. This bound is obtained by assuming that the excitation-number distribution has support only on the two nearest integer number sectors, which minimizes the variance for a fixed $N_e$. The close agreement between $\Delta N_e$ and its bound near the minima indicates that the system approaches the optimal charge-noise limit in these regions, while the deviations away from the bound at intermediate $n_L$ reflect the presence of additional particle-hole excitations beyond the minimal configuration.

\subsection{Role of spatial extension of voltage drop}

\begin{figure}
    \centering
    \includegraphics[width=0.5\textwidth]{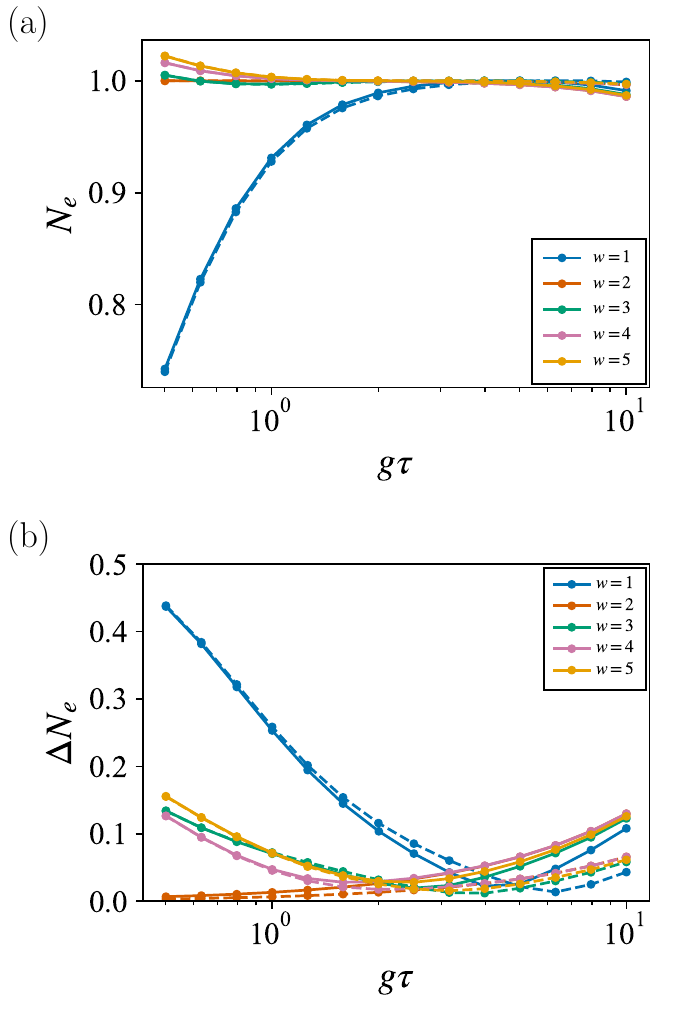}
    \caption{
    {(a)} Excitation number $N_e$ and
    {(b)} standard deviation $\Delta N_e$ as functions of pulse width $\tau$ with linear voltage drop over $w=1$ (blue), $w=2$ (red), $w=3$ (green), $w=4$ (violet), and $w=5$ (gold) sites. Trends in $N_e$ are nearly identical between $n=500$ (solid) and $n=1000$ (dash) systems, with the latter exhibiting smaller $\Delta N_e$. Width of the voltage drop plays little role in leviton creation with all widths converging once $\tau$ reaches near-optimal values. The parameters used are $gT_\text{cut}=n/8$, $n_L=1$, and $\nu=0.5$. }
    \label{fig:NeDelNe_dropWidth}
\end{figure}

Figure~\ref{fig:NeDelNe_dropWidth} shows the dependence of the excitation number and its variance on the pulse width $\tau$ for a linear voltage drop across different number of site bonds $w$ for a 500-site chain. A clear distinction emerges between the standard case, the case $w=2$, and broader voltage drops. For $w=1$, $N_e$ exhibits a strong dependence on $\tau$, increasing from small values at short pulse widths and approaching $N_e \approx 1$ only for $\tau \gtrsim 1$ for the range shown here, as seen previously in \Figref{fig:NeDelNeVStau_chainlength}. In contrast, for $w \geq 2$, the excitation number remains close to unity over a broad range of $\tau$, including the short-pulse regime. This indicates that wider voltage drops can more efficiently generate a single-particle excitation, even when the pulse duration is short (since, as we have already seen, in the small-$\tau$-limit, the unitary generated by the $w=1$ pulse approaches the identity). 

The behavior of the variance, shown in \Figref{fig:NeDelNe_dropWidth}(b), further highlights this distinction. For $w=1$, $\Delta N_e$ exhibits a well-separated maximum at small $\tau$, reflecting the generation of multiple particle-hole excitations. This demonstrates that voltage drops over multiple sites can improve the quality of single-particle excitations for shorter pulse durations compared to a voltage drop across a single site.
For $w \neq 2$, we see a monotonic decrease in the variance from small $\tau$ until a local minima forms. The local minima for $w\geq3$ are shifted towards smaller $\tau$ compared to the minima seen for $w=1$.  

The clear outlier in \Figref{fig:NeDelNe_dropWidth} is $w = 2$. In the short-pulse regime ($\tau \ll 1$) the voltage effectively acts as a local phase transformation on the lattice, and a voltage drop across two sites leads in the limit $\tau\to 0$ to a phase $\pi$ imparted on the site experiencing half of the applied voltage (and phases 0 or $2\pi$ on all other sites), which can be shown (at half filling and for even $n$) to correspond to an exact single-particle excitation in the excited subspace, yielding $N_e = 1$ and $\Delta N_e = 0$. However, the resulting state is not a leviton as it lacks localization in plane-wave space and contains contributions from both positive- and negative-$k$ plane waves. For example, although the state obtained with the short $w=2$-pulse at $g\tau \approx 0.50$ has $N_e \approx 1.000$ and $\Delta N_e \approx 0.0068$, 
it still differs markedly from a good leviton, as it contains a substantial contribution to the left of the voltage drop (about $11\%$ of the excited-state weight remains on the left half of the chain, compared with only about $2.5\%$ for the near-optimal $w=1$ leviton).

For larger widths $w \geq 3$, this special cancellation is lost, and although $N_e$ remains close to unity, the standard deviation increases with decreasing $\tau$, indicating the presence of additional particle-hole pairs.

Overall, \Figref{fig:NeDelNe_dropWidth} demonstrates that while increasing the width of the voltage pulse drop can significantly improve the purity of the excitation in the fast-pulse regime, this improvement is not observed at intermediate and large $\tau$. The convergence of $N_e$ and $\Delta N_e$ near $\tau_\text{opt}$ suggest that the assumption of a single-bond voltage drop is not necessary in generating levitonic excitations in finite systems. Furthermore, the improvement seen at short $\tau$ does not necessarily correspond to the generation of a well-defined leviton. This highlights an important distinction between minimizing charge noise and achieving the desired dynamical and spectral properties of the excitation.

\subsection{Role of different pulse shape}

\begin{figure}
    \centering
    \includegraphics[width=0.5\textwidth]{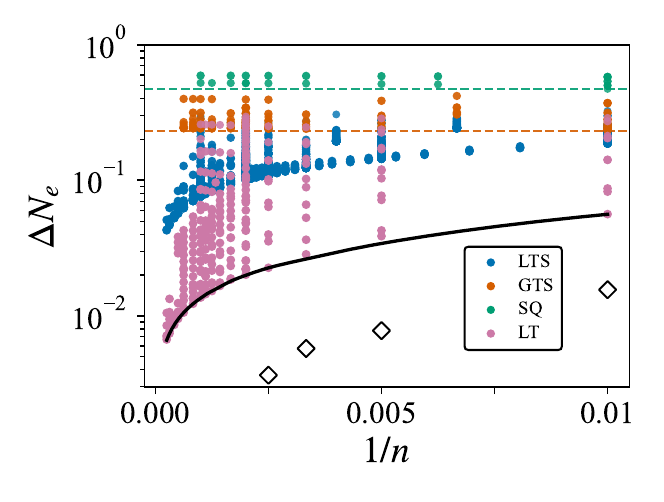}
    \caption{\label{fig:Ne_lengthscaling} Standard deviation in the number of excitations, for data satisfying $0.9 \le N_e \le 1.1$, as a function of inverse chain length for different voltage profiles. The dashed horizontal lines indicate the minimum displayed values of $\Delta N_e$ for the shifted truncated Gaussian (GTS) and square (SQ) datasets. The black solid curve shows the truncated Lorentzian (LT) optimal branch $\Delta N_e(\tau_\text{opt})$ for $gT_\mathrm{cut}=n/8$. Diamonds mark LT calculations from the larger-cutoff family $gT_\mathrm{cut}=2n$. Among the pulse families shown, the LT and LTS pulses are the only ones that show improvement in $\Delta N_e$ with increasing chain length. The LT and LTS parameters are  $n_L=1$, $\nu=0.5$, $w=1$.}
    \label{fig:DelNeScatter}
\end{figure}

In the continuum theory, the Lorentzian pulse shape plays a distinguished role in creating minimal noise excitations \cite{Levitov1996}. One may wonder to what extent this special role persists in the discrete setting in which finite-size effects modify the Fermi sea on which the excitation is created and truncation causes the pulse shape to be only approximately Lorentzian  (in that it includes discontinuities due to the finite time window and integrates to less than $2\pi$). 

Figure~\ref{fig:DelNeScatter} presents the standard deviation $\Delta N_e$ as a function of inverse chain length for a set of simulations with $0.9\leq N_e \leq 1.1$. The simulations span many different pulse parameters and include the truncated Lorentzian (LT), shifted truncated Lorentzian (LTS), shifted truncated Gaussian (GTS) and square pulse (SQ) profiles discussed in \Appref{app:pulseshape}. The shifted pulse shapes are chosen such that they are continuous and integrate to $2\pi$. Each point corresponds to a distinct calculation. The scatter therefore includes many sub-optimal configurations, but all have been simulated at half-filling ($\nu=0.5$) with a voltage drop across a single bond ($w=1$) at the center of the chain. The Lorentzian data points show pulses corresponding to $gT_\text{cut}=n/8$ only. The smallest recorded $\Delta N_e$ for the Gaussian and square pulse profiles are indicated by orange and green dashed guidelines respectively. The $\tau=\tau_\text{opt}$ simulations from \Figref{fig:NeDelNeVStau_chainlength}(d) have been interpolated to form the solid black guideline for the LT pulse. 

Several trends are immediately apparent. First, only for the Lorentzian pulse profiles are the minimal achievable standard deviations decreasing with increasing system size (decreasing inverse chain length). This is consistent with convergence toward the infinite-chain limit in which $\Delta N_e$ is expected to approach zero. Second, the shape of the pulse plays a crucial role in determining how closely this limit can be approached. The LT (pink) systematically define the lower boundary of the scatter, demonstrating their superior ability to suppress additional particle-hole excitations. In contrast, GTS (orange) and SQ (green) pulses typically yield larger variances, reflecting less efficient cancellation of unwanted excitations -- as does the rescaled and LTS: these pulse modifications markedly degrade its performance with respect to $\Delta N_e$. We note that the $\tau=\tau_\text{opt}$ guideline does not represent a general threshold. These simulations were performed using $gT_\text{cut}=n/8$ and by increasing the pulse duration systematic improvement is seen across all chain lengths as indicated by the diamond symbols (showing the result for $gT_\text{cut}=2n$). This improvement for $gT_\text{cut}>n/4$ shows that neither self-interference nor the presence of excitations at the position of the voltage drop necessarily degrade the leviton generation.

The distinction between pulse shapes becomes less pronounced for shorter chains ($n\lesssim 100$, not shown), particularly between the LTS, GTS, and SQ pulses. In this regime, the shift and renormalization of the LTS pulse compensate for the advantage of the Lorentzian profile, while the LT pulse still shows roughly an order of magnitude reduction in $\Delta N_e$. However, as the system size increases, the advantage of Lorentzian pulses becomes increasingly clear, exhibiting the importance of its optimal temporal structure for leviton formation in the large-system limit.

\section{Physical timescales and requirements}
\label{sec:discussion}

The finite-lattice effects identified above have direct implications for possible experimental realizations. In the TB model, the relevant dynamical scale is set by the hopping amplitude $g$, with characteristic times measured in units of $\hbar/g$ and characteristic velocities of order $2ga/\hbar$ near half filling. For graphene, the interatomic atomic spacing $a\simeq 1.4 \, \mathrm{\AA}$ and nearest-neighbor $p_z$-orbital hopping parameter of the order $g=3\,\mathrm{eV}$, corresponds to a velocity of order $10^{6}\, \mathrm{m/s}$ at the Fermi level \cite{CaGuPe.09.electronicpropertiesof}. A wave packet moving at this speed traverses a 10 nm structure in roughly 10 fs. This places atomically precise graphene structures in a challenging regime: their large bandwidth supports rapid coherent propagation, but the same bandwidth compresses the experimentally relevant time and length scales.

The corresponding voltage and field scales are also severe for an atomistic graphene-based implementation. For a truncated Lorentzian pulse with $T_\text{cut}\gg\tau$, the peak voltage scale is set approximately by the inverse pulse width. In dimensionless TB units, pulses with $g\tau$ of order unity therefore require peak voltage amplitudes comparable to the electronic bandwidth scale $g/e$. For graphene, this corresponds to voltage drops on the order of volts across atomic distances, implying electric fields of order $10^{10}\, \mathrm{V/m}$ if the voltage drop occurs over a single bond. Such fields should be viewed as an indication that the simplest single-bond voltage-drop model is not directly realistic. In experimental graphene leviton devices, by contrast, Lorentzian voltage pulses are applied to extended quantum Hall edge channels over micron-scale geometries, with much smaller applied voltages and periodic pulse trains \cite{Assouline2023}. Similarly, the original GaAs leviton experiments employed long mesoscopic channels rather than atomically short conductors \cite{Dubois2013}. The present model should therefore be interpreted as a microscopic lattice proxy for understanding how discreteness, finite bandwidth, and pulse truncation affect leviton formation, rather than as a direct prescription for applying an atomic-scale voltage drop in graphene.

An alternative route is to engineer the graphene band structure itself. In GNRs, the relevant group velocity can be substantially reduced relative to pristine graphene. For example, flattening of electronic bands near the Fermi energy has been achieved in GNR superlattices \cite{RiVeCa.18.Topologicalbandengineering,GrWaYa.18.Engineeringrobusttopological,RiVeJi.20.Inducingmetallicitygraphene}. Since the leviton propagation time scales inversely with the characteristic velocity, such band engineering can shift the experimentally relevant pulse durations toward longer times without abandoning graphene-based materials altogether. From the perspective of the present TB description, this corresponds to reducing the effective bandwidth and characteristic propagation velocities governing the dynamics. In this sense, the experimentally relevant parameter is not the microscopic carbon-carbon hopping amplitude but the effective low-energy bandwidth and dispersion experienced by the levitonic excitation.

Moir\'e graphene systems provide an intermediate regime between atomically defined GNRs and gate-defined quantum-dot arrays. In these materials, the relevant lattice scale is not the carbon--carbon spacing, but the moir\'e period, which is typically on the order of $10$--$15\, \mathrm{nm}$ near the magic angle. A single effective site in a low-energy TB description can therefore represent a collective electronic state spread over many carbon atoms within one moir\'e unit cell. At the same time, the electronic bandwidth can be reduced from the graphene scale of several eVs to the meV scale by twist-angle engineering, producing flat or narrow moir\'e bands whose properties are tunable by twist angle, displacement field, dielectric environment, and lattice relaxation \cite{Bistritzer2011, Cao2018, Andrei2020, Liu2020}. For levitonic wave-packet generation, this combination is attractive. The enlarged lattice constant relaxes the spatial-resolution requirement while the reduced effective hopping slows the dynamics into a more experimentally accessible time window. For example, an effective hopping in the range of $1$--$10\,\mathrm{meV}$ and a moir\'e period of order $10\,\mathrm{nm}$ would correspond to characteristic propagation velocities several orders of magnitude below those of atomic graphene, moving the relevant pulse durations from the fs regime towards the ps regime. Moir\'e platforms therefore offer a promising compromise between the structural precision of graphene-based materials and the slower, highly tunable dynamics of quantum-dot arrays. However, their usefulness for leviton physics will depend on whether sufficiently coherent 1D channels or guided paths can be engineered within the moir\'e landscape, and on whether interaction in the narrow bands can be controlled well enough for the non-interacting TB picture used here to remain a useful starting point.

Gate-defined quantum-dot arrays provide a complementary regime in which the relevant lattice parameters are much more tunable. In such systems, the effective hopping amplitude depends sensitively on interdot separation and electrostatic gate voltages. For example, values around $g=40\, \mu\mathrm{eV}$ have been considered for quantum-dot arrays with lattice spacings of order $160\,\mathrm{nm}$ \cite{Knoerzer2022}. These parameters reduce the characteristic velocity by many orders of magnitude compared with graphene and move the relevant pulse durations into the ps-to-ns regime. A 100-dot array with this spacing has a length of order $16\, \mu\mathrm{m}$, and the corresponding traversal time is naturally in the sub-ns to few-ns range, depending on the precise operating point and filling. This makes quantum-dot arrays more favorable form the perspective of pulse generation and electric-field strength. At the same time, reducing $g$ too far is not automatically beneficial, since levitonic dynamics must remain fast compared with dephasing and charge-noise timescales. The design problem is therefore one of balancing controllability against coherence: small hopping amplitudes relax the pulse-generation requirements, but also increase the time over which the wave packet is exposed to environmental noise.

Another platform to realize finite-size levitons are cold fermionic atoms in optical lattices \cite{Bloch2008, Tarruell2018}. These would provide single-site control and read-out as well as in-situ tunability of hopping, on-site energy and interactions, allowing, in particular, to realize effectively non-interacting fermions as discussed here as well as a gradual increase in interactions to study their effect on leviton formation and scattering. Here, very different time and energy scales are probed, with $g$ on the order of $\sim10^{-c}\mu$eV, i.e., frequency scales of kHz \cite{Xu2025a}. 

These estimates clarify the physical meaning of the optimal-pulse regime found in the finite-chain simulations. The requirement that the Lorentzian profile be sufficiently resolved by the lattice favors longer pulses, while the need to avoid excessive truncation and dephasing favors shorter propagation times. In graphene-based structures, the large hopping amplitude pushes the compromise toward ultrafast control and extended geometries. In quantum-dot arrays, the same compromise can be shifted into a slower and more experimentally accessible regime by tuning the effective hopping. Thus, although the microscopic parameters differ substantially between material platforms, the dimensionless constraints identified in the TB model provide a useful guide for assessing where clean leviton-like excitations may be feasible.

\section{Conclusions}
\label{sec:conclusions}

We have investigated the generation and characterization of levitonic excitations in finite 1D TB chains. Starting from the non-interacting fermionic lattice driven by time-dependent voltage pulses, we evolved the SPDM and projected the resulting state onto the initially empty subspace of the static Hamiltonian. This excited-space SPDM provides a compact description of the generated particle excitation above the Fermi sea and allows the leviton quality to be quantified in terms of the average excitation number $N_e$, and its standard deviation $\Delta N_e$.

The time-resolved analysis of the excited-space density matrix further shows that leviton formation is a coherent many-mode process. During the first half of the pulse, population is transferred into several modes above the Fermi sea, producing a fragmented excitation with appreciable impurity. During the second half of the Lorentzian pulse, the subleading eigenvalues are suppressed while the dominant eigenvector evolves into a localized, predominantly right-moving wave packet with the energy-space structure expected for a Lorentzian excitation. 

Our results show that leviton formation in a finite lattice is controlled by a competition between pulse propagation and pulse truncation. Pulses that are too short are not well resolved by the discrete chain and fail to generate clean Lorentzian-like excitations, while pulses that are too broad are strongly affected by the finite cutoff and finite system size. Between these limits, we identify an optimal intermediate regime in which a Lorentzian voltage pulse generates a near-integer excitation with substantially reduced number fluctuations. The optimal pulse width increases with chain length, but does so sublinearly over the range of system sizes studied, reflecting the tension between resolving the central part of the pulse and retaining enough of its long Lorentzian tails. The residual value of $\Delta N_e$ decreases with increasing chain length, consistent with convergence toward the continuum minimal-excitation limit, although the computationally accessible finite systems remain visibly distant from the ideal limit.

We also find that minima in $\Delta N_e$ occur at integer values of the pulse parameter $n_L$ in agreement with continuum models. Band filling affects the spatial structure and propagation of the generated excitation through the filling dependence of the Fermi velocity and the curvature of the TB dispersion. In particular, half filling is distinguished by particle-hole symmetry and by the vanishing of the leading band-curvature correction at the Fermi point, whereas away from half filling the leviton and antileviton profiles become asymmetric and propagate differently.

A comparison between Lorentzian, Gaussian, and square pulses confirms the special role of the Lorentzian pulse shape. For sufficiently large chains, Lorentzian pulses systematically define the lowest-noise boundary among the pulse shapes considered and show the clearest improvement with increasing system size. Gaussian and square pulses can be tuned to produce approximately one excitation on average, but they do not suppress number fluctuations with the same efficiency, and in fact show a slight increase in $\Delta N_e$ with increasing chain length, consistent with expected diverging noise for non-Lorentzian pulses in infinite systems.

We further examined the role of the spatial profile of the voltage drop. Extending the voltage drop over multiple bonds can improve the excitation-number statistics in the fast-pulse regime, but this improvement does not necessarily  imply the creation of a well-defined leviton. In particular, special short-pulse cases can yield small $\Delta N_e$ while producing states with poor spatial and plane-wave index localization. This highlights an important distinction between minimizing charge noise and generating a propagating wave packet with the desired spectral and dynamical properties. Consistently, we find that $\Delta N_e$ is a useful primary diagnostic, especially near the optimal Lorentzian regime, 
but not sufficient to identify good levitons in all edge cases. 

Finally, we discussed the physical timescales associated with possible material platforms. Atomically precise GNRs provide a natural microscopic realization of TB physics, but their large hopping amplitudes imply ultrafast propagation and correspondingly demanding pulse-generation and electric-field requirements. Moir\'e systems and gate-defined quantum-dot arrays offer complementary regimes in which the effective lattice spacing and hopping amplitudes can be tuned, potentially shifting levitonic dynamics into more experimentally accessible time windows.

Overall, this work provides a microscopic lattice-level characterization of leviton generation beyond the ideal continuum limit. By identifying how finite bandwidth, lattice dispersion, filling, pulse truncation, and linear voltage geometry affect the formation of leviton-like excitations, we establish practical criteria for using TB models as proxies for engineered electron-quantum optics platforms. These results provide a basis for future studies of leviton propagation in more realistic GNR networks, moir\'e channels, and quantum-dot arrays, where disorder, interactions, spin structure, and multi-terminal geometries can be incorporated into the same density-matrix framework.

\section*{Acknowledgments}
We thank Ricardo Ortiz and Guido Burkard for stimulating discussions in the early phases of this project.
The research was funded by the Department of Education of the Basque Government through PIBA\_2023\_1\_0021 (TENINT), by Agencia Estatal de Investigación MCIN/AEI/10.13039/501100011033 through Proyectos de Generación de Conocimiento PID2023-146694NB-I00 (GRAFIQ), and by the IKUR Strategy under the collaboration agreement between Ikerbasque Foundation and DIPC on behalf of the Department of Education of the Basque Government with in project QT13: DREAMS.

\appendix
\section{Additional Pulse Shapes}
\label{app:pulseshape}
Here we provide the expressions for the three alternative voltage profiles considered in the main text.

\subsection{Shifted truncated Lorentzian pulse}
The shifted truncated Lorentzian (LTS) pulse, we use the form  
\begin{equation}\label{eq:LTS}
V^\text{LTS}(t)=
A^\text{LTS}_{n_L}\left[
\dfrac{\tau}{t^2+\tau^2}
-\dfrac{\tau}{T_{\rm cut}^2+\tau^2}
\right] \theta (T_{\rm cut} - |t|),
\end{equation}
with $\theta(x)=1$ for $x\geq0$ and $0$ otherwise and amplitude
\begin{equation}\label{eq:PulseNorm}
A^\text{LTS}_{n_L}=
\frac{2\pi n_L}{
2\arctan\!\left(\frac{T_{\rm cut}}{\tau}\right)
-\frac{2\tau T_{\rm cut}}{T_{\rm cut}^2+\tau^2}}.
\end{equation}
This normalization ensures that the total accumulated phase is equal to $2\pi n_L$. In the continuum limit, this condition corresponds to the injection of exactly $n_L$ levitons. In the finite lattice studied here, $n_L$ is treated as a tunable parameter that characterizes the pulse amplitude.

\subsection{Shifted truncated Gaussian pulse}
The shifted truncated Gaussian (GTS) voltage profile is defined as
\begin{equation}
  V^{\rm GTS}(t)=
  A_{n_L}^{\rm GTS}
  \left[
  e^{-t^2/\tau^2}
  -
  e^{-T_{\rm cut}^2/\tau^2}
  \right] \theta(T_{\rm cut} - |t|), 
\end{equation}
with prefactor
\begin{equation}
  A_{n_L}^{\rm GTS}
  =
  \frac{2\pi n_L}{
  \sqrt{\pi}\,\tau\,\operatorname{erf}\!
  \left(T_{\rm cut}/\tau\right)
  -2T_{\rm cut}e^{-T_{\rm cut}^2/\tau^2}
  }.
\end{equation}
Again, this construction ensures $V^{\rm GTS}(\pm T_{\rm cut})=0$.

\subsection{Square pulse}
The square pulse (SQ) is compactly supported and therefore does not require any shift as for the other pulses. Defining
\begin{equation}
  t_*=\min\!\left(\frac{\tau}{2},T_{\rm
  cut}\right),
\end{equation}
we write
\begin{equation}
  V^{\rm SQ}(t)=
  A_{n_L}^{\rm SQ}\theta(t_* - |t|),
\end{equation}
with
\begin{equation}
  A_{n_L}^{\rm SQ}=\frac{2\pi n_L}{2t_*}
  =\frac{\pi n_L}{t_*}.
\end{equation}
For \(T_{\rm cut}\ge \tau/2\), this simplifies to
\begin{equation}
  V^{\rm SQ}(t)=
  \dfrac{2\pi n_L}{\tau} \theta(\tau/2 -  |t|).
\end{equation}

\section{Derivation of \Eqref{eqn:varNe}}\label{app:varNe}

Single particle observables can be written in the form $A=\sum_{i,j}a_{i,j}c^\dagger_i c_j$ on the single-particle space. The expectation value of such an observable in the state $\varrho$ is 
\begin{equation}
    \langle A\rangle_{\varrho} = \sum_{ij} a_{ij}\varrho_{ij}\equiv \tr{a \varrho^T}.
\end{equation}
Its variance can be obtained (for Gaussian states) using the factorizing property:
\begin{align} 
\nonumber\langle A^2\rangle_{\varrho} &= \sum_{ijkl} a_{ij}a_{kl}\langle c_i^\dagger c_j c_k^\dagger c_l\rangle_{\varrho},\\
\nonumber&=\sum_{i} a_{ii}^2  \langle c_i^\dagger c_i\rangle_{\varrho} + 
\sideset{}{'}\sum_{ijkl} a_{ij}a_{kl} \langle c_i^\dagger c_j c_k^\dagger c_l\rangle_{\varrho}\\
\nonumber&=\sum_{i} a_{ii}^2  \varrho_{ii} \\
\nonumber &\quad + 
\sideset{}{'}\sum_{ijkl} a_{ij}a_{kl} \left(\langle c_i^\dagger c_j\rangle_{\varrho} \langle c_k^\dagger c_l\rangle_{\varrho}-\langle c_i^\dagger c_l\rangle_{\varrho} \langle c_k^\dagger c_j\rangle_{\varrho}\right)\\
\nonumber&=\sum_{i} a_{ii}^2  \varrho_{ii} + 
\sum_{ijkl} a_{ij}a_{kl} \left(\varrho_{ij} \varrho_{kl}-\varrho_{il} \varrho_{kj}\right)\\
\label{eq:SPvariance} &= \tr{a^2 \varrho^T} + \tr{a \varrho^T}^2 - \tr{a \varrho^T a \varrho^T}.
\end{align}
In the second step the primed sum is only over $(ijkl)$ that are not all the same as these have been treated in the first sum. In the third line we used the factorization property of Gaussian states~\cite{BLS94} and also that for our systems there is no pairing ($\langle c_i c_j\rangle=0)$.  
Higher moments can be computed similarly, requiring longer sums.

\section{Numerical integration of the dynamics}\label{app:numerics}
The dynamics of the $n$-site chain governed by the
time-dependent single-particle Hamiltonian $H(t)=H_0+H_1(t)$ is governed by the time-evolution operator
\begin{equation}
  \label{eq:2}
  U(t_f,t_i) = \mathcal{T}e^{-i\int_{t_i}^t ds H(s)},
\end{equation}
where $\mathcal{T}$ is the time-ordering operator. We use the Magnus expansion \cite{Blanes2009} to approximate 
\begin{eqnarray}
    U(t_l+\Delta t,t_l) &\approx &\exp[-i\Delta t H_{\text{eff}, l}],\\
    -i\Delta t H_{\text{eff}, l} &=& \sum_{m\geq0} \Omega^{(l)}_m
\end{eqnarray} 
for sufficiently small time steps $\Delta t$ with an time-independent generator. To avoid the overhead implied by matrix exponentiation, which maintains unitarity exactly, we write $\exp[A]=\exp[A/2](\exp[A^\dagger/2])^{-1}$ (for antihermitian $A=-i\Delta t H_{\text{eff}, l}$) and approximate each factor by 
    \begin{equation}\label{appeq:expA}
    \exp[A/2] \approx 1+A/2+\frac{1}{2!}(A/2)^{2}+\dots+\frac{1}{n!}(A/2)^{n},
\end{equation}
which is correct to order
$A^{n}$ if $n$ is even and to order $A^{n+1}$ if $n$ is odd.

Accuracy can be increased either by choosing $\Delta t$ sufficiently small or by going to high order in the Magnus expansion. The former implies more matrix multiplications due to many time steps, the latter allows for larger time steps (since the operators $\Omega_{(m), l}$ are of order 
$\Delta t_{l}^{m}$ or higher \cite{Ikeda2023}), but leads to increasingly cumbersome expressions for the generators, which result from $m$-fold nested 
integrals over the interval $[t_{l},t_{l}+\Delta t]$ of nested commutators of $H_0$ and $H_1(t)$.  
 For our calculations, we choose Magnus expansion up to the $m=2$ term, thus neglecting terms of order $\Delta t^5$ \cite{Ikeda2023}. Consequently, we then also neglect all terms of order higher than $\Delta t^4$ from \Eqref{appeq:expA}.

\begin{widetext}
To make the steps clearer, consider $\exp[A/2]$ with $A=\Omega=\sum_{m\geq1}\Omega_{m}$:
\begin{align*}
  e^{A/2}&\approx\left( 1 + \Omega/2 +\frac{1}{2!}(\Omega/2)^{2}
    +\frac{1}{3!}(\Omega/2)^{3}+\frac{1}{4!}(\Omega/2)^{4}\right)\\
       &\approx \left( 1 + \frac{1}{2}\Omega_{1}+ 
         \frac{1}{2!2^{2}}\Omega_{1}^{2}+\frac{1}{2}\Omega_{2}+
         \frac{1}{3!2^3}\Omega_{1}^{3}  + \frac{1}{2!2^2}(\Omega_{1}\Omega_{2}+\Omega_{2}\Omega_{1}) + \frac{1}{4!2^4}\Omega_{1}^{4}\right),
\end{align*}
where we have omitted terms of order higher than $\Delta t^{4}$ since we have already accepted an error of 5th order by discarding $\Omega_{3}$
\footnote{Note that $\Omega_{2}$ is of order $\Delta t^{3}$ while $\Omega_{3}$ and $\Omega_{4}$ are of order 
$\Delta t^{5}$.}.
\end{widetext}

An important practical simplification comes from the time-dependence of $H(t)$ being given by a single term, a fixed operator $H_1$ being multiplied by a time-dependent factor $V(t)$. This reduces the number of commutators and time-integrals needed to compute the $\Omega_{l}$. Following \cite{Ikeda2023}, we find
\begin{align}
  \Omega_1 &= \int_{t_{l}}^{t_{l+1}}ds (H_0 +
             V(s)H_1)\nonumber\\
   \Omega_2 &= \frac{1}{2} \int_{t_l}^{t_{l+1}}ds_1\int_{t_l}^{s_1} ds_2
             [H_0+V(s_1)H_1,H_0+V(s_2)H_1]\nonumber\\
  &= \frac{1}{2} \int_{t_l}^{t_{l+1}}ds_1\int_{t_l}^{s_1} ds_2 
    \left(V(s_2)-V(s_1)\right) [H_0,H_1].
\end{align}
The required time-integrals are evaluated using the Taylor series expansion of $V(s)$ around the center of the interval in question to the appropriate order. This has the advantage that now all integrals can be computed analytically in terms of the Taylor coefficients (which have simple expression in terms of the parameters $\tau$ and $T_\text{cut}$ for all shapes considered).

\section{Fast excitations using $w=2$: clean, non-levitonic single-particle excitations}
We noticed a remarkable behavior of the excitations created by the width $w=2$ voltage pulse in the limit of short $\tau$: it generates almost ideal single-particle excitations, i.e., states with $N_e=1$ and $\Delta N_e\approx0$ and does so for all system sizes, so apparently outperforming the standard $w=1$ pulses.

This behavior can be easily understood in the fast limit $g\tau\ll 1$. In this limit, the voltage pules simply imparts a phase on every site. This phase depends on the site index and is $\Phi=2\pi$ for all sites to the left of the voltage drop, $\Phi=0$ to the right of the drop and $\Phi=\pi$ at the site $j_0$ which experiences half the pulse strength. Applied to the half-filled initial state, this produces an exact pure single-excitation state in the $\tau\to0$ limit.

The initial state is represented by the SPDM $\varrho_0=P_\mathrm{occ}=\sum_{l=1}^{n_F}\ketbra{e_l}{e_l}$, where the sum is over the lowest $n_F=n/2$ energy eigenstates of $H_0$. 
The fast pulse transforms $\rho_0$ to $\rho=FP_0F^\dagger$, where $F=\text{diag}(1,\dots,1,-1,1,\dots1)$, with $-1$ at site $j_0=n/2+1$. Each energy eigenvector is transformed by $F$ as
\begin{equation}
F\ket{e_l} = \ket{e_l}-2e_{l,j_0}\ket{j_0},
\end{equation}
where $e_{l,j_0}$ is the amplitude of the $l$th energy eigenvector at site $j_0$. 
This implies 
\begin{align}\label{eq:Pf}
    \varrho &=F\varrho_0F^\dagger\\
    &= P_0 -2\sum_l \left(e_{l,j_0}^* \ketbra{j_0}{e_l} + \mathrm{h.c.}\right) + 4\sum_l |e_{l,j_0}|^2\ketbra{j_0}{j_0}.
\end{align}
We are interested in the e-SPDM $\varrho_e=P_e\varrho P_e$ at the end of the fast pulse. The matrix elements of $\rho_e$ in the energy eigenbasis are:
\begin{align*}
[\varrho_e]_{rs} &= \bra{e_r}FP_0F^\dagger \ket{e_s}\\
&=4 e_{r,j_0}^*e_{s,j_0}\sum_{l=1}^{n_F} |e_{l,j_0}|^2,
\end{align*}
since only the third term in Eq.~(\ref{eq:Pf}) contributes. 

For our chain, we have $e_{l,j}=\sqrt{\frac{2}{N+1}}\sin(\frac{\pi l j}{N+1})$ and we can show that for $n_F=n/2$
\begin{widetext}
\begin{equation*}
    \sum_{l=1}^{n_F} |e_{l,j_0}|^2 = \frac{2}{n+1}\sum_l \sin^2[\frac{\pi lj_0}{n+1}]
    =\frac{1}{4(n+1)}\left(n+1 - \frac{\sin(\pi j_0)}{\sin[\pi j_0/(n+1)]}\right)
    = \frac{1}{2},
\end{equation*}
\end{widetext}
since $j_0$ is an integer. (Here, $j_0$ needs not be in the center of the chain. This says that at half filling, the state represented by a single particle at site $j_0$ is with equal probability 1/2 in the initially filled or empty space.)
So we have
\begin{align}
[\varrho_e]_{rs} &=2 e_{r,j_0}^*e_{s,j_0}.
\end{align}

$N_e$ can now be easily computed as the trace of $\varrho_e$, which is 
\begin{equation}
4\times\Big(\sum_{l\leq n/2} |e_{l,j_0}|^2\Big)\Big(\sum_{l<n/2} |e_{l,j_0}|^2\Big) = 1.
\end{equation}
The calculation of the variance can be simplified by noting that free (Gaussian) states are pure only if they have a sharp particle number. Hence $N_e=1$ and purity 1 implies $\Delta N_e=0$. That $\varrho_e$ is pure follows from
\begin{align*}
    \tr{\varrho_e^2} &= \sum_{k,l>n/2} [\varrho_e]_{kl}[\varrho_e]_{lk}\\
    &=4 \sum_{k,l}e_{k,j_0}^*e_{l,j_0}e_{l,j_0}^*e_{k,j_0}\\
    &=4\Big(\sum_{k>n/2} |e_{k,j_0}|^2\Big)^2\\
    &=1 \equiv \tr{\varrho_e},
\end{align*}
so $\varrho_e$ is indeed pure and has exactly on particle in one mode.

But in the small-$\tau$ limit the mode that is occupied is not a nice leviton-like excitation: it contains both positive and negative momenta and has amplitudes both to the left and to the right of the voltage drop. One can see that, e.g., in the form of a vanishing bond current or $\langle K_e\rangle$ dropping to zero (at finite variance).

Along the same lines, one readily sees that the fast $w=1$ pulse leaves the system unchanged (all site acquire either phase $2\pi$ or $0$. We have not analyzed the $w>2$ case in detail. But now several sites acquire a non-trivial phase and numerical calculation show no special behavior in the fast limit. The e-SPDM in this case is mixed.

\bibliography{references}

\end{document}